\documentclass[twocolumn,unsortedaddress]{revtex4-1}
\usepackage{epsfig,pstricks,graphicx}
\usepackage{amsmath,amssymb}
\usepackage[mathscr]{eucal}
\usepackage{color}
\usepackage{comment}
\usepackage[normalem]{ulem}

\def\CC{{\rm\kern.24em \vrule width.04em height1.46ex depth-.07ex
\kern-.30em C}}
\def\RR{{\rm
         \vrule width.04em height1.58ex depth-.0ex
         \kern-.04em R}}
\def\P{{\rm I\kern-.25em P}}
\def\id{{\rm 1\kern-.22em l}}

\newcommand{\bra}[1]{\left\langle #1 \right |}
\newcommand{\ket}[1]{\left | #1 \right\rangle}

\newcommand{\oost}{\frac{1}{\sqrt{2}}}
\newcommand{\rme}{\ensuremath{\mathrm{e}}}
\newcommand{\rmi}{\ensuremath{\mathrm{i}}}
\newcommand{\rmd}{\ensuremath{\mathrm{d}}}

\newcommand{\tr}{\operatorname{tr}}

\begin{document}
\title{
Partial transpose as a direct link between concurrence and negativity
      }

\author{Christopher Eltschka}
\affiliation{Institut f\"ur Theoretische Physik, 
         Universit\"at Regensburg, D-93040 Regensburg, Germany}

\author{G\'eza T\'oth}
\affiliation{Department of Theoretical Physics, 
             University of the Basque Country UPV/EHU, 
             P.O.\ Box 644, E-48080 Bilbao, Spain
            }
\affiliation{IKERBASQUE, Basque Foundation for Science, E-48013 Bilbao, Spain}
\affiliation{Wigner Research Centre for Physics, Hungarian
Academy of Sciences, P.O. Box 49, H-1525 Budapest, Hungary}

\author{Jens Siewert}
\affiliation{Department of Physical Chemistry,
             University of the Basque Country UPV/EHU, 
             P.O.\ Box 644, E-48080 Bilbao, Spain}
\affiliation{IKERBASQUE, Basque Foundation for Science, E-48013 Bilbao, Spain}

\begin{abstract}
Detection of entanglement in bipartite states is a fundamental
task in quantum information. The first method to verify entanglement
in mixed states was the partial-transpose criterion. Subsequently,
numerous quantifiers for bipartite entanglement were introduced,
among them concurrence and negativity. Surprisingly,
these quantities are often treated as distinct or independent of each other.
The aim of this contribution is to highlight the close relations between
these concepts, to show the connections between seemingly
independent results, and to present various estimates for the mixed-state
concurrence within the same framework.
\end{abstract}
\maketitle
\date{\today}


\section{Introduction}
In quantitative entanglement
theory, considerable effort has been spent in developing many
different entanglement 
measures~\cite{Horodecki2009,GuehneToth2009,EltschkaReview2014}, 
while much less is known regarding the
relations between these measures and, in particular,
their connection to the resources they quantify.
This has lead to a situation that parts of quantitative 
entanglement theory have the fame of a certain arbitrariness, 
if not a lack of meaning.
As opposed to this, we believe that---given a well-defined entanglement
measure---there is a physical resource (defined through a protocol)
that is quantified by this measure. 

At present, there are only few
well-established links between entanglement-related resources and their 
quantifiers~\cite{Bennett1996a,Bennett1996b,Horodecki3-1999,Grondalski2002,Pezze2009,Hyllus2012,Toth2012}. 
However, many more mathematical ways of characterizing
and quantifying entanglement  are known than corresponding protocols using
that entanglement.
Therefore we think it is important to investigate and reveal
the relations between different concepts, their possible common origins
and essential differences, in order to introduce more structure 
in the world of entanglement measures where it is possible.

The subject of this article exemplifies the coexistence
and apparent independence of different concepts in entanglement
characterization. The partial-transpose 
criterion~\cite{Peres1996,Horodecki1996} provided the first possibility to
detect entanglement in arbitrary mixed states. Later, numerous
tools based on the partial transpose were developed, such as
decomposable entanglement witnesses~\cite{Lewenstein2000}, 
negativity (and logarithmic negativity) as an entanglement
measure~\cite{Zyczkowski1998,Eisert1999,Eisert2001,Vidal2002}, combinations of
the latter in detection of multipartite 
entanglement~\cite{Jungnitsch2011,Hofmann2014}, and others.
On the other hand, the concurrence was first introduced by Bennett
{\em et al.}~\cite{Bennett1996b} as an auxiliary tool to compute the
entanglement of formation for Bell-diagonal two-qubit
states and then developed
further by Wootters and co-workers~\cite{Hill1997,Wootters1998,CKW2000} who
established concurrence as an entanglement measure in its own right.
Subsequently, generalizations to the higher-dimensional case $d\times d$
($d>2$) as well as for multipartite systems 
(e.g., Refs.~\cite{Rungta2001,Albeverio2001,Badziag2002,Rungta2003,Akhtarshenas2005,Love2006,Ma2011,Huber2012})
were proposed.

There have been comparative studies of concurrence and negativity
(e.g., Refs.~\cite{Eisert1999,Audenaert2001,Zyczkowski2001,Miranowicz2004,Grudka2004,Fei2005},
however, as far as we can see  they continue to exist in separate 
research lines. Therefore, we find it useful to present a discussion
showing that both negativity and concurrence can be directly related
to the partial transpose, and that both essentially determine
the same type of entanglement deriving 
from the Schmidt rank of a Bell state
which results in various mathematical relations linking all these quantities.

Our paper is organized as follows.
In Sec.\ II, we introduce
the most important concepts and notation.
In Sec.\ III, we discuss the links between partial transpose,
concurrence, and negativity for pure states. Finally, in Sec.\ IV, we extend this
discussion to mixed states. In particular we study
a family of symmetric bipartite mixed states---the axisymmetric states---for 
which the quantitative concepts of interest can be derived exactly
thus providing both an illuminating illustration as well as a 
powerful tool for further investigation.

\section{Definitions}

Throughout this article we study bipartite quantum systems 
with $d$-dimensional local parties (often termed $d\times d$ systems).
For their pure states 
$\psi\in \mathcal{H}=\mathcal{H}_A\otimes \mathcal{H}_B=
\mathbb{C}^d\times \mathbb{C}^d$. Given orthonormal bases
$\{\ket{j}_A\}$, $\{\ket{k}_B\}$ for the two parties a state $\psi$
can be written
\begin{equation}
      \ket{\psi}\ =\ \sum_{j,k=1}^d \psi_{jk}\ket{j}_A\otimes\ket{k}_B
                \ \equiv\ \sum_{j,k=1}^d \psi_{jk}\ket{jk}
\ \ .
\label{eq:purestate}
\end{equation}
The mixed states $\rho$ are bounded positive operators acting on $\mathcal{H}$
and can be represented as convex combinations of pure-state projectors
$\pi_{\psi}\equiv \ket{\psi}\!\bra{\psi}$ (with $\tr\pi_{\psi}=1$)
\begin{equation}
      \rho\ =\ \sum_j\ p_j\ \pi_{\psi} \ \ ,
\label{eq:decomp}
\end{equation}
where $p_j\geqq 0\ ,\ \sum_j p_j =1$.
Importantly, such decomposition of a mixed state is not unique, that is,
there are infinitely many pure-state ensembles representing a given  
state~\cite{HJW1993}.

For each pure state $\psi$, there is a so-called 
{\em Schmidt decomposition}~(see, e.g., Ref.~\cite{Preskill1998})
\begin{equation}
      \ket{\psi}\ =\ \sum_{j=1}^{r(\psi)} \sqrt{\lambda_{j}}\
                     \ket{\alpha_j\beta_j}
\label{eq:schmidtrank}
\end{equation}
with $\lambda_j\geqq 0$ and $r(\psi)\leqq d$.
The local Schmidt bases $\{\ket{\alpha_j}_A\}$, $\{\ket{\beta_j}_B\}$ can
be obtained be suitable local unitary transformations from given
local bases $\{\ket{k}_A\}$, $\{\ket{l}_B\}$. The {\em Schmidt rank}
$r(\psi)$ is the number of nonvanishing Schmidt coefficients $\lambda_j$.

For mixed states, the generalization of the Schmidt rank is the
{\em Schmidt number}~\cite{Terhal2000}, 
defined as the smallest maximum Schmidt rank
occurring in any decomposition $\{(p_j,\psi_j)\}$ of $\rho$
\begin{equation}
      r(\rho)\ =\ \min_{\{(p_j,\psi_j)\}} \max_j\ r(\psi_j)\ \ .
\label{eq:schmidtnumber}
\end{equation}

Given a state $\rho$, we may trace out one of the parties
and obtain the reduced state of the other party $\rho_A\equiv \tr_B\rho$
(and analogously for $\rho_B$). A well-known quantifier for the
entanglement of the pure state $\psi$ is the {\em concurrence}
\begin{equation}
      C(\psi)\ =\ \sqrt{2(1-\tr\rho_A^2)}\ =\ \sqrt{2(1-\tr\rho_B^2)}
\ \ .
\label{eq:defconc}
\end{equation}
Note that sometimes the factor 2 is replaced by $d/(d-1)$, however, 
this difference in normalization is not essential.
Interestingly it was shown~\cite{Albeverio2001,Badziag2002} that 
\begin{align}
    C(\psi) & \ =\ \sqrt{\sum_{jklm} |\psi_{jm}\psi_{lk}-\psi_{jk}\psi_{lm}|^2}
\nonumber\\
    &\ =\ \sqrt{4 \sum_{j<l,k<m} |\psi_{jm}\psi_{lk}-\psi_{jk}\psi_{lm}|^2} \ \ ,
\label{eq:concvecconc}
\end{align}
where $c_{jklm}\equiv\psi_{jk}\psi_{lm}-\psi_{jm}\psi_{lk}$
are the {\em concurrence vector componenents}~\cite{Badziag2002,Audenaert2001}.

For mixed states $\rho$
the concurrence is given by the minimum average concurrence
taken over all decompositions of $\rho$, the so-called 
convex roof~\cite{Uhlmann2010}
\begin{equation}
      C(\rho)\ =\ \min_{\{(p_j,\psi_j)\}} \sum_j p_j C(\psi_j)
\ \ .
\label{eq:convexroof}
\end{equation}
The convex roof is notoriously hard to evaluate, therefore
it is difficult to determine whether or not an arbitrary
state is entangled.

The {\em partial transpose} of a bipartite state $\rho$ (with respect
to party $B$) is given by
\begin{equation}
      \rho^{T_B}\ =\ \left[\sum_{jklm} 
                     \rho_{jk,lm} \ket{jk}\!\bra{lm}\right]^{T_B}
       \equiv\ \sum_{jklm} 
                     \rho_{jk,lm} \ket{jm}\!\bra{lk}
\ \ .
\label{eq:PT}
\end{equation}
The partial transpose indicates entanglement of $\rho$
if $\rho^{T_B}$ has negative eigenvalues. The corresponding
quantifier is the {\em negativity}~\cite{Zyczkowski1998,Eisert1999,Eisert2001,Vidal2002}
\begin{equation}
   \mathcal{N}(\rho)\ =\ \frac{1}{2}\left( ||\rho^{T_B}||_1-1 \right)\ ,
\label{eq:defneg}
\end{equation}
where
$||M||_1\equiv\tr\sqrt{M^{\dagger}M}$ is the trace norm of the matrix $M$.
The huge advantage of the negativity is that it can easily be computed
for any mixed state, however, at the price that entanglement in states
with a positive partial transpose (PPT) is not detected.

A related quantity which does detect PPT entanglement but, again, is hard
to compute is the {\em convex-roof extended negativity}~\cite{Lee2003,Zhang2013}
\begin{equation}
   \mathcal{N}^{\mathrm{CREN}}(\rho)\ =\ 
                  \min_{\{(p_j,\psi_j)\}} \sum_j p_j \mathcal{N}(\psi_j)
\ \ .
\label{eq:defCREN}
\end{equation}
It is the largest convex function that conincides with $\mathcal{N}(\psi)$
on the pure states, that is, $\mathcal{N}^{\mathrm{CREN}}(\rho)\geqq
\mathcal{N}(\rho)$.

An important property of concurrence and negativity is that
they are both invariant under local unitary transformations.
The Schmidt rank does not change under arbitrary 
invertible local operations.

\section{Pure states}

The relation between pure-state concurrence and the partial transpose
was studied before (e.g., \cite{Audenaert2001,Fei2005}) but we find
it important to make it very explicit here, because this will directly
connect to the negativity and show what is actually quantified
by these measures. Moreover, we discuss how concurrence and negativity
can be viewed as $\ell_p $ norms of the concurrence vector.

\subsection{Partial transpose and concurrence}
Assume the pure state $\phi$ is a tensor product
\begin{align}
    \ket{\phi} = &  \sum_{jk}\ \phi_{jk}\ \ket{jk}
\nonumber \\
               = &  \ket{a}\otimes\ket{b} = \sum_{jk}\ a_j b_k\ \ket{jk}
\end{align}
so that the corresponding projector can be written
\begin{align}
   \pi_{\phi} = & \ \sum_{jklm} \phi_{jk}\phi_{lm}^*\ket{jk}\!\bra{lm}
\nonumber\\ 
              = & \ \sum_{jklm} a_j b_k a^*_l b^*_m \ket{jk}\!\bra{lm}
\ \ .
\end{align}
The matrix elements of the partial transpose of $\pi_{\phi}$ instead read
\begin{align}
   \left(\pi_{\phi}^{T_B}\right)_{jk,lm} =   \phi_{jm}\phi_{lk}^*
              =   a_j b_k^* a^*_l b_m 
\ \ .
\end{align}
Therefore, for any product state $\phi$ we have
\begin{align}
    \left|\left(\pi_{\phi}^{T_B}\right)_{jk,lm}\right|^2- &
    \left(\pi_{\phi}^{T_B}\right)_{jk,jk}\left(\pi_{\phi}^{T_B}\right)_{lm,lm}
\nonumber\\
         =\ & |\phi_{jm}\phi_{lk}|^2-|\phi_{jk}\phi_{km}|^2
\nonumber\\
         =\ & |\phi_{jm}\phi_{lk}-\phi_{jk}\phi_{lm}|^2\ =\ 0\ \ .
\label{eq:condPPT}
\end{align}
A state that does not fulfill condition~\eqref{eq:condPPT} cannot
be a product state. Since violation of Eq.~\eqref{eq:condPPT} may
occur for any combination of level pairs $(j,l)$ for party $A$ and $(k,m)$
for $B$, we define
\begin{equation}
    \tilde{C}(\psi)^2\ \equiv\ 
               \sum_{jklm} \left|\psi_{jm}\psi_{lk}-\psi_{jk}\psi_{lm}\right|^2 
\label{eq:concveclength}
\end{equation}
as a quantifier for the violation of the product-state condition 
for the state $\psi$. By comparing 
Eqs.~\eqref{eq:concvecconc} and \eqref{eq:concveclength}
we see that $\tilde{C}(\psi)$ coincides with the $d\times d$ concurrence
$C(\psi)$. That is, {\em for a pure state $\psi$ 
                         the squared concurrence is simply a measure for the
                         total violation of  the PPT-type 
                         condition~\eqref{eq:condPPT}.
                    }

Because of the local unitary invariance of $C(\psi)$ 
(see Sec.~\ref{sec:concneg})
it suffices
to compute the concurrence for the Schmidt decomposition 
$\ket{\psi}=\sum \sqrt{\lambda_j}\ket{\alpha_j\beta_j}$
of $\psi$, one obtains
\begin{align}
   C(\psi)\ = & \ \sqrt{2\sum_{jk} \left|\sqrt{\lambda_j\lambda_k}\delta_{jk}
                               -\sqrt{\lambda_j\lambda_k}\right|^2}
\nonumber\\ = & \ 2\ \sqrt{\sum_{j< k} \lambda_j\lambda_k}\ \ ,
\label{eq:concSchmidt}
\end{align}
that is, the concurrence is identical with the case $k=2$ of 
the $k$-concurrence defined
by Gour~\cite{Gour2005}, the $k$th elementary symmetric function of the 
Schmidt coefficients taken to a power so that it is homogeneous of
degree 2 in the state coefficients (note the different normalization
of $C_2(\psi)$ in Ref~\cite{Gour2005}). 

Consider now the
(maximally entangled) Bell state of 
rank $r$
\begin{equation}
       \ket{\Phi_r}\ =\ \frac{1}{\sqrt{r}}\sum_{j=1}^r \ \ket{jj}\ \ .
\label{eq:bell}
\end{equation}
The corresponding concurrence equals
\begin{equation}
      C(\Phi_r)\ =\ \sqrt{\frac{2(r-1)}{r}}\ \ .
\label{eq:concBell}
\end{equation}
We see that the concurrence grows monotonically with the
Schmidt rank of $\Phi_r$. For rank-$r$ states which are not maximally entangled
the concurrence clearly is smaller than $C(\Phi_r)$, 
in a sense it attributes an `effective rank' 
$r_{\mathrm{eff}}=\frac{1}{1-\frac12 C^2}<r$ to the state.

\subsection{Negativity}
\label{sec:Neg}

Again, because of the local unitary invariance, 
we can compute the negativity 
$\mathcal{N}(\psi)$ from the Schmidt decomposition 
$\ket{\psi}=\sum\sqrt{\lambda_j}\ket{\alpha_j\beta_j}$
\begin{equation}
    \mathcal{N}(\psi)= \sum_{j<k} \sqrt{\lambda_j\lambda_k}
\ \ .
\label{eq:negSchmidt}
\end{equation}
In particular, we find for the Bell states $\Phi_r$
\begin{equation}
    \mathcal{N}(\Phi_r)\ =\ \frac12\left( 2\frac{r(r-1)}{2}\frac{1}{r}
                                  \right)
                       \ =\ \frac{r-1}{2}
\ \ .
\label{eq:negBell}
\end{equation}
That is, in analogy with the concurrence the negativity `counts'
the Schmidt rank, if the state is maximally entangled. The word `counting'
can be taken literally here due to the linear dependence of 
$\mathcal{N}(\Phi_r)$ on $r$ (cf.~Ref.~\cite{ES2013}).

Thus, we see that both concurrence and negativity quantify the
Schmidt rank, albeit in a mathematically different manner.
If a state $\psi$ is not maximally entangled,
both measures attribute a kind of `effective rank' to it
which is smaller than that of the maximally entangled state
locally equivalent to $\psi$ [that is, equivalent under
stochastic local operations and communication (SLOCC)].

From Eqs.~\eqref{eq:concBell} and \eqref{eq:negBell} it is evident
that the negativity gives equal weight to each Schmidt rank
increment while the concurrence favors increments
at lower Schmidt ranks. It is not difficult to track down the origin of
this difference by comparing the squared equations \eqref{eq:concSchmidt} and
\eqref{eq:negSchmidt}. 
The squared concurrence contains only the products of two
different Schmidt coefficients
while the squared negativity has contributions also from products of up to four
Schmidt coefficients.

Moreover, the negativity keeps increasing linearly with the Schmidt rank
while the concurrence converges to a finite value.  This means that
adding more dimensions to a state which already has  high Schmidt
rank does practically not augment the concurrence.
This hints at the fact that concurrence and negativity, 
both being related to the Schmidt rank of the state, 
quantify
qualitatively different resources:
The resource quantified by the concurrence is present to a high degree
already in a state with relatively low Schmidt rank, 
and can be improved beyond that only marginally. On the other hand, 
the negativity can grow without a limit on increasing the Schmidt rank, 
and this  should apply also to the
corresponding resource.

\subsection{Concurrence, negativity, and $\ell_p$ norms}
\label{sec:concneg}

From Eq.~\eqref{eq:concvecconc} and also form the derivation of 
condition \eqref{eq:condPPT} leading to the total violation of the PPT
condition, we see that the concurrence
formally looks like the length of a Euclidean vector, i.e.,
it can be regarded as the $\ell_2$ norm of the concurrence vector.
However, in our derivation of condition \eqref{eq:condPPT} it was
by no means necessary to use the square of 
$|\psi_{jm}\psi_{lk}-\psi_{jk}\psi_{lm}|$. The last line of
Eq.~\eqref{eq:condPPT} is correct also without squaring it.
Hence, we could have introduced a total violation of the PPT condition
also as
\begin{equation}
           \tilde{\mathcal{N}}(\psi)\ \equiv\ \frac14
           \sum_{jklm} \left|\psi_{jm}\psi_{lk}-\psi_{jk}\psi_{lm}\right| 
\ \ .
\label{eq:negPPT}
\end{equation}
Now, comparing Eq.~\eqref{eq:negPPT} with the negativity calculated from
the Schmidt decomposition, Eq.~\eqref{eq:negSchmidt},
it is tempting to conclude
that the pure-state negativity actually {\em equals}
the $\ell_1$ norm of the concurrence
vector~\cite{Ma2014}, i.e., 
$\mathcal{N}(\psi)= \tilde{\mathcal{N}}(\psi)$. 
Unfortunately this is not correct in general.
The reason is that the right-hand side in Eq.~\eqref{eq:negPPT} may
increase under local unitaries, and thus cannot represent an
entanglement monotone.

To put it in different words: The negativity of a pure state $\psi$
equals the $\ell_1$ norm \eqref{eq:negPPT} of the concurrence vector
$c_{jklm}=(\psi_{jk}\psi_{lm}-\psi_{jm}\psi_{lk})$ 
if $\psi$ is given in the Schmidt basis. Then, this $\ell_1$ norm assumes
its {\em minimum}
\begin{equation}
           \mathcal{N}(\psi)\ =\ \min_{\text{local\ bases}}\tilde{\mathcal{N}}
                                     (\psi) \ \ ,
\label{eq:negmin}
\end{equation}
while for other local bases it is larger. 
Clearly, since the {\em minimum of the} $\ell_1$ {\em norm} \eqref{eq:negmin}
equals the negativity
it is an entanglement monotone, however, the $\ell_1$ norm written 
in a different basis is not (it is not even invariant under local unitaries).

Now we will
prove  that $\tilde{\mathcal{N}}(\psi)$ is minimized
for  $\psi$ given in the Schmidt basis, 
Eq.~\eqref{eq:schmidtrank}.
To this end, consider local unitaries $U$ and $V$ applied to the 
parties of $\psi$ written in the Schmidt basis
\begin{equation}
\tilde{\psi}_{ab}\ =\ \sum_{mn} U_{am}V_{bn}\psi_{mn} \ \equiv\
                      \sum_{mn} U_{am}V_{bn}\sqrt{\lambda_m}\delta_{mn}
\end{equation}
and use this in Eq.~\eqref{eq:negPPT} to express $\tilde{\mathcal{N}}(\tilde{\psi})$  
as
\begin{align}
     \tilde{\mathcal{N}}&(\tilde{\psi})  = \frac14
            \sum_{aa'bb'}  |\tilde{\psi}_{ab}\tilde{\psi}_{a'b'} -
                           \tilde{\psi}_{ab'}\tilde{\psi}_{a'b}|
\nonumber\\
         &= \frac14 \sum_{aa'bb'} \left|\sum_{mn} U_{am}U_{a'n} 
            \left(V_{bm}V_{b'n}-V_{bn}V_{b'm}\right)
            \sqrt{\lambda_m\lambda_n}\right|\ \ .
\nonumber
\end{align}
In order to proceed, we note that
\begin{align}
          \sum_{aa'bb'} \left|\sum_{mn} U_{am}U_{a'n} 
            \left(V_{bm}V_{b'n}-V_{bn}V_{b'm}\right)
            {x_{mn}}\right|^2
\nonumber  
\\  =\ 2 \sum_{mn} \left|x_{mn}\right|^2 ,
\label{eq:unitconc}
\end{align}
which is easily seen by expanding $|y|^2=y y^*$.
This relation also implies
local unitary invariance of the concurrence.

By substituting 
$x_{mn}=(\sqrt{\lambda_m\lambda_n}-1)$ 
in Eq.~\eqref{eq:unitconc} and 
applying the triangle inequalities $|a-b|\geqq ||a|-|b||
\geqq |a|-|b|$ we obtain
\begin{align}
2 & \sum_{m\neq n}  \sqrt{\lambda_m\lambda_n}
\nonumber\\
\leqq & \sum_{aa'bb'}
         \left|\sum_{mn}
      |U_{am}U_{a'n}\left(V_{bm}V_{b'n} -V_{bn}V_{b'm}\right)
         \sqrt{\lambda_m\lambda_n}\right|\times
\nonumber\\
      & \ \ \ \times
         \left|\sum_{mn}
      |U_{am}U_{a'n}\left(V_{bm}V_{b'n} -V_{bn}V_{b'm}\right)
         \right|
\ \ .
\end{align}
The last factor on the right-hand side
is $\leqq 1$, by virtue of
the Cauchy-Schwarz inequality and the normalization of columns
of unitary matrices. Hence we have
\begin{align}
\mathcal{N}(\psi)\ =\  
\mathcal{N}(\tilde{\psi})\ =\  
\sum_{m>n} \sqrt{\lambda_m\lambda_n} \ \leqq\ 
     \tilde{\mathcal{N}}(\tilde{\psi}) \ ,
\end{align}
where, as defined above, $\psi$  is a state
given in the Schmidt basis, while $ \tilde{\psi}$  is obtained from $\psi$ 
by applying local unitaries. 
 
Finally, it is easy to construct an example showing
that indeed $\tilde{\mathcal{N}}$ can increase under local unitaries,
consider, e.g., $\Phi_3$ and apply a Hadamard transform in the subspace
$\{\ket{1},\ket{2}\}$. This concludes the proof. 

\subsection{Inequalities for concurrence and negativity}

By using the results of the previous section, a number of interesting
inequalities connecting concurrence, negativity and Schmidt rank can
be proven.

We have already mentioned that for a pure state 
$\psi\in \mathbb{C}^d\times\mathbb{C}^d$ of Schmidt rank $r$ 
\begin{align}
        C(\psi)\ & \leqq\ \sqrt{\frac{2(r-1)}{r}}\ \ ,\nonumber
\\
        \mathcal{N}(\psi)\ & \leqq\ \frac{r-1}{2}\ \ .\nonumber
\end{align}
Furthermore, we see that
\begin{align}
        2\mathcal{N}(\psi)\  \geqq\ C(\psi)\ \geqq\ \
               2\sqrt{\frac{2}{r(r-1)}}\ \mathcal{N}(\psi)
        \ \ .
\label{eq:negpure-ineq2}
\end{align}
The first of these inequalities can be readily deduced from
Eqs.~\eqref{eq:concSchmidt} and \eqref{eq:negSchmidt} while
the second is a consequence of the fact that the quadratic 
is larger than the arithmetic mean. Obviously, 
for pure states of two qubits the
negativity equals the concurrence divided by two.

\begin{figure}[t!]
  \centering
 \includegraphics[width=.97\linewidth]{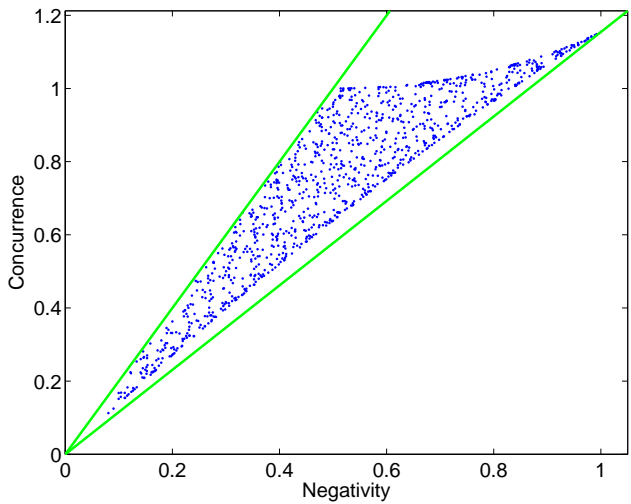}
  \caption{(color online).
Concurrence vs.\ negativity for 1000 random pure states
\cite{random}
 (blue dots)
           with $d=4$ and $r\leqq 3$.
           The green lines represent the upper and lower bounds
           in Eq.~\eqref{eq:negpure-ineq2}.
          }
\label{fig:bounds}
\end{figure}

In Fig.~\ref{fig:bounds}, we illustrate the bounds of Eq.~\eqref{eq:negpure-ineq2}
by plotting the concurrence and negativity for many randomly
chosen pure states. Clearly, those linear estimates are not the best ones possible.
In fact, the evident (curved) boundaries for the concurrence values can be obtained
by analytically maximizing/minimizing the concurrence for given negativity and
rank of the state.

%

\section{Mixed states}

The estimation of the Schmidt number and the concurrence
for arbitrary mixed states is an important problem, both for
theory and experiment.
By combining the results for pure states and some recent
ideas~\cite{ES2013,Ma2011} one obtains an interesting toolbox
for practical applications.

   \subsection{Concurrence, negativity, and Schmidt number}

The great advantage of the negativity is that it can be evaluated
also for arbitrary mixed states, as opposed to the Schmidt number or
the concurrence. We will now discuss some relations connecting the negativity
to the other two quantites.

It has been shown only recently~\cite{ES2013} that
the negativity can actually be used as a lower bound  on the Schmidt number.
Let us assume that  we are given a mixed state $\rho$ of a $d\times d$ system
and a decomposition $\{( p_j,\psi_j)\}$ 
that optimizes the Schmidt number $r(\rho)$. Then we have
%
\begin{equation*}
   \mathcal{N}(\rho) \leqq \sum_j p_j \mathcal{N}(\psi_j)
                     \leqq \sum_j p_j \frac{r(\rho)-1}{2}
                     \leqq \frac{r(\rho)-1}{2}
\end{equation*}
from which follows 
\begin{equation}
  r(\rho) \ \geqq\ 2\mathcal{N}(\rho)+1\ \ .
\label{eq:negdimbound}
\end{equation}
Here we have used Eq.~\eqref{eq:negBell} and the convexity
of negativity.
An analogous inequality can be derived for the concurrence
\begin{equation}
   C(\rho) \leqq \sum_j p_j C(\psi_j)
                     \leqq \sqrt{\frac{2(r(\rho)-1)}{r(\rho)}} \ \ ,
\end{equation}
from which we arrive at
\begin{equation}
  r(\rho) \ \geqq\ \frac{1}{1-\frac12 C(\rho)^2}\ \ .
\end{equation}
This relation is essentially different from Eq.~\eqref{eq:negdimbound}.
On the one hand, finding a useful lower bound for the
concurrence might be more difficult than calculating the negativity. 
On the other hand, for PPT-entangled states 
Eq.~\eqref{eq:negdimbound} is not useful.

We can also find inequalities for the concurrence and the negativity.
From the first inequality in Eq.~\eqref{eq:negpure-ineq2}
it follows that
\begin{equation}
   2\mathcal{N}^{\mathrm{CREN}}(\rho)\ \geqq\ C(\rho)\ \ ,
\label{eq:CRENvsC}
\end{equation}
while from the second inequality of Eq.~\eqref{eq:negpure-ineq2}
we get for an optimal decomposition $\{(p_j,\psi_j)\}$ of the concurrence 
(cf.~\cite{Audenaert2001,Fei2005})
\begin{align}
        C(\rho)\  =\ \sum_j p_j C(\psi_j)  
                &\geqq\ 
               2\sum_j\ p_j\sqrt{\frac{2}{r_j(r_j-1)}} \mathcal{N}(\psi_j)
\nonumber\\
                & \geqq\ 
               2\sqrt{\frac{2}{r(r-1)}}\ \mathcal{N}^{\mathrm{CREN}}(\rho)
\nonumber\\
                & \geqq\ 
               2\sqrt{\frac{2}{r(r-1)}}\ \mathcal{N}(\rho)
\ ,
\label{eq:CvsN}
\end{align}
where $r_j=r(\psi_j)$ and $r=\max\ r_j$.
If $r$ is not known it can be replaced by the
dimension $d$. 

We mention that
the two-qubit concurrence divided by two  {\em equals} the
convex-roof extended negativity $\mathcal{N}^{\mathrm{CREN}}$ 
[cf.~Eq.~\eqref{eq:negpure-ineq2}], and the 
negativity $\mathcal{N}$ is a lower bound to the latter.
Therefore, the $2\times 2$ negativity never exceeds
the concurrence divided by two, as noted in~Ref.~\cite{Audenaert2001}.
In contrast, for higher local dimension $d>2$
the negativity may be larger 
than half the concurrence (e.g., Bell states), but it may also be smaller 
(e.g., PPT-entangled states).

   \subsection{Systematic lower bounds for concurrence}
   \label{sec:lowerbounds}

Already from the discussion in the preceding section it can be seen
that it is desirable to have systematic ways for estimating the
mixed-state concurrence. However, this has proven difficult over
the years. Only recently, an elegant method was devised by Huber
and co-workers (based on earlier ideas~\cite{Guhne2010,Huber2010})
to lower bound concurrence-type entanglement 
measures, see, e.g., Ref.~\cite{Ma2011,Huber2012}. While in those references
the focus is on multipartite states, we apply it here in the simpler
case of bipartite states. 

The method proceeds in two steps. First, we estimate the pure-state
concurrence directly from Eq.~\eqref{eq:concvecconc}. Subsequently
we show that the resulting inequality applies to mixed
states. We select a set $\mathcal{M}$ of $\mu$ pairs $\{jk,lm\}$
$(j<l,k<m)$ 
and estimate the corresponding terms in Eq.~\eqref{eq:concvecconc}
using the triangle inequality and the inequality between arithmetic
and quadratic mean 
\begin{align}
   C(\psi)\ & \geqq\ \frac{2}{\sqrt{\mu}} \sum_{jklm\in \mathcal{M}}
                                   |\psi_{jk}\psi_{lm}-\psi_{jm}\psi_{lk}|
\nonumber\\
    & \geqq\ \frac{2}{\sqrt{\mu}} \sum_{jklm\in \mathcal{M}}
                       |\psi_{jk}\psi_{lm}| -\sqrt{|\psi_{jm}|^2|\psi_{lk}|^2}
\ \ .
\label{eq:huber1}
\end{align}
The convex-roof construction for the concurrence and
the convexity of the functions on the right-hand
side of Eq.~\eqref{eq:huber1} guarantee that we can replace
all state components by the corresponding density matrix elements
so that
\begin{align}
   C(\rho)\  
     \geqq\ \frac{2}{\sqrt{\mu}} \sum_{jklm\in \mathcal{M}}
                       |\rho_{jk,lm}| -\sqrt{\rho_{jm,jm}\rho_{lk,lk}}
\ \ .
\label{eq:huber2}
\end{align}
Because of the local unitary invariance of the concurrence,
one can maximize this lower bound simply by changing local bases.

A nice application of this inequality results if we choose 
$\mathcal{M}$ such that it specifies the off-diagonal matrix elements
$\rho_{jj,kk}$ ($j<k$) of the Bell-state projector $\pi_{\Phi_d}$,
i.e., $\mu=\frac12 d(d-1)$. For simplicity we replace the
square root terms by $\frac12(\rho_{jk,jk}+\rho_{kj,kj})\geqq
\sqrt{\rho_{jk,jk}\rho_{kj,kj}}$ and obtain
\begin{subequations}
\begin{align}
   C(\rho)  
   & \geqq  \sqrt{\frac{2}{d(d-1)}} \sum_{j<k}
                       \left(\rho_{jj,kk}+\rho_{kk,jj}
               -\rho_{jk,jk}-\rho_{kj,kj}\right)
\nonumber\\
    \geqq & \sqrt{\frac{2}{d(d-1)}}\left[ -1+\sum_{j<k}
                       (\rho_{jj,kk}+\rho_{kk,jj})
               +\sum_{j}\rho_{jj,jj}\right]
\nonumber
\\
    \geqq &\ \ \sqrt{\frac{2d}{d-1}} 
                \ \Big[\ \langle \Phi_d | \rho | \Phi_d\rangle
                                   \     -\ \frac{1}{d}\ \Big]
\label{eq:huber3}
\\
    \geqq &\ \ \sqrt{\frac{2d}{d-1}} \tr \left(\rho
                \ \Big[\ \ket{\Phi_d}\!\bra{\Phi_d}
                                   \     -\ \frac{1}{d}\id_{d^2} \Big]
                                         \right)\ \ ,
\label{eq:huber4}
\end{align}
\end{subequations}
which is  a concurrence estimate from 
the optimal Schmidt number witness~\cite{Sanpera2001} (for Schmidt number 2).

We can even improve this bound by optimization
over local unitaries.
This way we encounter another well-known quantity, namely the
{\em fully entangled fraction} $\mathcal{F}$~\cite{Bennett1996b}
\begin{equation}
   \mathcal{F}(\rho)\ =\ \max_{U_A,U_B} \bra{\Phi_d}(U_A\otimes U_B) \rho
                                       (U_A\otimes U_B)^{\dagger} \ket{\Phi_d}
\label{eq:FEF}
\end{equation}
and Eq.~\eqref{eq:huber3} then reads
\begin{equation}
    C(\rho)\ \geqq \  \max \left(0,\sqrt{\frac{2d}{d-1}} 
                \ \Big[\ \mathcal{F}(\rho)
                                   \     -\ \frac{1}{d}\ \Big]\right)
\ \ ,
\label{eq:FEFconc}
\end{equation}
a result found in Ref.~\cite{Fei2010}.

   \subsection{Axisymmetric states}

We conclude our survey by considering a nontrivial 
family of mixed states for which the quantitative concepts 
we have discussed can be evaluated exactly for all finite dimensions $d$.
This family is called {\em axisymmetric states}~\cite{ES2013}. 
In $d$ dimensions it includes all those states that
have the same symmetries as the Bell state $\Phi_d$, Eq.~\eqref{eq:bell},
namely
\begin{itemize}
  \item[(i)] permutation symmetry of the two qudits,
  \item[(ii)]  symmetry with respect to simultaneously exchanging two
        levels for both qudits, e.g., $\ket{1}_A \leftrightarrow \ket{2}_A$
        and  $\ket{1}_B \leftrightarrow \ket{2}_B$,
  \item[(iii)]  simultaneous (local) phase  rotations of the form
\begin{equation}
          \;\;\;\; \;\; \;\;\;\;V(\varphi_1,\varphi_2,\ldots,\varphi_{n-1})=
          \rme^{\rmi\sum \varphi_j \mathfrak{g}_j}\otimes
          \rme^{-\rmi\sum \varphi_j \mathfrak{g}_j}\ \ ,
\label{eq:phase}
\end{equation}
\end{itemize}
      where $\mathfrak{g}_j$ are the $(d-1)$ diagonal generators
      of SU($d$).
Note that the period of the phase angles $\varphi_j$ depends on the
normalization of the generators $\mathfrak{g}_j$. For axisymmetric states,
qubit permutation symmetry is implied by the requirements
(ii) and (iii).

After discussing the symmetries of axissymmetric states, 
we will show how to parametrize them.
In any dimension $d$ the $d\times d$ axisymmetric
states are parametrized by two real numbers. This can be
seen as follows. 
The phase rotation symmetry eliminates all off-diagonal
components which are not of the form $\rho_{jj,kk}$. 
Qudit permutation and simultaneous level exchange symmetry are possible
only if all off-diagonal elements are real and equal (one parameter)
and there are only two different types of diagonal elements
($\rho_{jk,jk}$ for $j=k$ and $j\neq k$) which give one more
parameter, due to the normalization constraint $\tr \rho =1$.
Based on the ideas above, we have the following parametrization
\begin{equation}
          \rho^{\mathrm{axi}}_{jj,jj} \ =\ \frac{1}{d^2} + a \  ,\ \ \ 
          \rho^{\mathrm{axi}}_{jk,jk} \ =\ \frac{1}{d^2}-\frac{a}{d-1}\ \ (j\neq
 k)
\label{eq:diagels}
\end{equation}
($j,k=1,\ldots,d$) and off-diagonal entries
\begin{equation}
          \rho^{\mathrm{axi}}_{jl,km} \ =\ \left\{ \begin{array}{ll}
                                                b \ \ \ & \mathrm{for}\ l=j\ ,\ 
m=k
                                                \\
                                                0 \ \ \ & \mathrm{otherwise}\ \ 
.
                                                \end{array}
                                         \right.
\label{eq:offdiagels}
\end{equation}

\begin{figure}[t!]
  \centering
 \includegraphics[width=.97\linewidth]{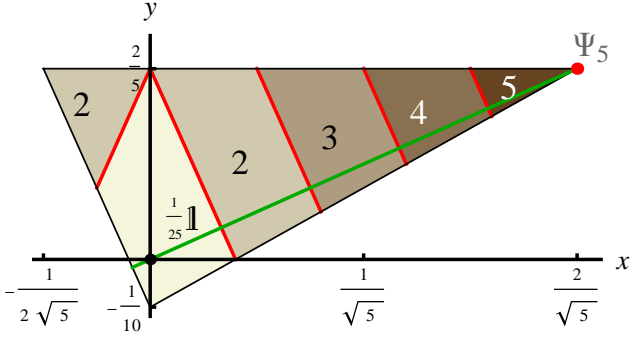}
  \caption{(color online). The  family of $d\times d$ axisymmetric states
           $\rho^{\text{axi}}$ for $d=5$.
              It is characterized
              by two real parameters
              $x$ and 
              $y$ describing the offdiagonal matrix elements
              and the asymmetry between the two types of
              diagonal elements, respectively 
              [see Eqs.~\eqref{eq:diagels}--\eqref{eq:xyscaling}].
              The upper right corner corresponds to the
              $\Phi_d$, the only
              pure state in the family. The completely mixed state 
              $\frac{1}{d^2}\id_{d^2}$ is located at the origin so that
              the isotropic states lie on the solid green line
              connecting the origin with the upper right corner.
              Note that this line is divided by the
              Schmidt number regions in $d$ parts of equal length.
               Hence the relative area of the separable states
              (compared to the total area of the triangle)
              tends to zero for $d\rightarrow\infty$
              so that for axisymmetric
              states of large local dimension $d$, separability
              is the peculiar feature, rather than entanglement,
              in agreement with the conclusion for the entire
              state space in Ref.~\cite{Zyczkowski1998}.
    }
\label{fig:triangle}
\end{figure}

Let us now determine the limits for the paramaters for physical states.
We are free to choose the length scales of $a$ and $b$ in such a way
that in a graphical representation the lengths are the same as
in state space, hence geometrical intuition can be directly
applied to the figures. Here the length in state space $D_{\mathrm{HS}}(A,B)$
is defined via the Hilbert-Schmidt scalar product 
$D_{\mathrm{HS}}^2(A,B)=\tr(A-B)(A-B)^{\dagger}$.
The appropriate scaling factors for the coordinates $x$ and $y$ are
\begin{equation}
      a\ =\ y\frac{\sqrt{d-1}}{d}\ \ ,\ \ \ b\ =\ \frac{x}{\sqrt{d(d-1)}}
\label{eq:xyscaling}
\end{equation}
from which we can compute the boundary of the axisymmetric states
\begin{subequations}
\begin{eqnarray}
\label{eq:range-y}
   -\frac{1}{d\sqrt{d-1}}\ & \leqq & \ y\ \leqq\ \frac{\sqrt{d-1}}{d} \ \ ,\\
\label{eq:range-x}
   -\frac{1}{\sqrt{d(d-1)}} \ & \leqq & \ x\ \leqq\ 
    \sqrt{\frac{d-1}{d}}    
\end{eqnarray}
\end{subequations}
as well as
\begin{equation} 
-\frac{1}{\sqrt{d}}\left(y+\frac{1}{d\sqrt{d-1}}\right)\leqq x
\leqq \frac{d-1}{\sqrt{d}}\left(y+\frac{1}{d\sqrt{d-1}}\right),
\label{eq:eqdreieck}
\end{equation}
i.e., we find a triangular shape for this family (cf.~Fig.~\ref{fig:triangle}).
In this parametrization, the completely mixed state $\frac{1}{d^2}\id_{d^2}$
is located at the origin while the Bell state $\Phi_d$ (the only pure
state in the family) appears in the upper right corner.
%
%

%
%
Based on the considerations above, 
we can conclude that the important isotropic states~\cite{Horodecki1999}
\begin{equation}
     \rho^{\mathrm{iso}}\ = \ p\ \ket{\Phi_d}\!\bra{\Phi_d}\ +\ 
                               \frac{1-p}{d^2}\ \id_{d^2}
\end{equation}
form a subfamily of the axisymmetric states. 
This is not a surprise because the symmetry group of the isotropic states 
is $U\otimes U^*$ where $U$ is an arbitrary local unitary and $U^*$ its complex
conjugate. The simultaneous phase rotations Eq.~\eqref{eq:phase}
form a subgroup of $U\otimes U^*$ (and isotropic states obey permutation
and level exchange symmetry), hence the isotropic states must be 
a subset of the axisymmetric family.

A particular advantage of state families defined via symmetries
is that it is possible to project an arbitrary state into the
families by averaging over the given symmetries~\cite{Vollbrecht2001}. 
Correspondingly, isotropic states can be obtained by averaging
(often termed {\em twirling}) over local unitaries $U$ 
\begin{equation}
\label{eq:projiso}
\mathbb{P}^{\mathrm{iso}}(\rho) = \int\rmd U \
                          (U\otimes U^*) \rho\, (U\otimes U^*)^\dagger \ \ ,
\end{equation}
while axisymmetric states arise from twirling over the
operations $\mathcal{V}$ including permutations and the local unitaries
$V$ in Eq.~\eqref{eq:phase} 
\begin{equation}
\label{eq:projaxi}
\mathbb{P}^{\mathrm{axi}}(\rho) = \int\rmd \mathcal{V} \
                          \mathcal{V} \rho\, \mathcal{V}^\dagger
\ \ .
\end{equation}
In these expressions the integral ``$\int \rmd X$'' denotes the average 
over the 
the corresponding symmetry group including the discrete symmetries.
We mention already at this point that these averages do not increase
the entanglement in the projection
$\mathbb{P}^{\mathrm{axi}}:\rho\rightarrow \rho^{\mathrm{axi}}$ (and analogously for
isotropic states) because neither permutations nor local unitaries
or mixing can increase entanglement.

The procedure for performing the average in Eq.~\eqref{eq:projaxi} 
in practice is easy:
Given an arbitrary $d\times d$ state $\rho$ the 
matrix elements of its projection $\rho^{\mathrm{axi}}(\rho)$
are
\begin{subequations}
\begin{align}
   \rho^{\mathrm{axi}}_{jj,jj} & \ =\ \frac{1}{d} \sum_m \rho_{mm,mm} \ ,
\\
   \rho^{\mathrm{axi}}_{jk,jk} & \ =\ \frac{1}{d(d-1)}\sum_{m\neq n} \rho_{mn,mn}\ \ (j\neq
 k)
\end{align}
\end{subequations}
with $j,k=1,\ldots,d$, and off-diagonal elements
\begin{subequations}
\begin{align}
          \rho^{\mathrm{axi}}_{jj,kk} & \ =\ 
                   \frac{1}{d(d-1)} \sum_{m>n} \left(\rho_{mm,nn}+\rho_{nn,mm}\right)  \ ,
\\       
          \rho^{\mathrm{axi}}_{jk,lm} & \ =\ 0\ \ \ \ \mathrm{for}\
                                                  k\neq j\ \mathrm{or}\ 
                                                  l\neq m
\ \ .
\label{eq:symmmatels}
\end{align}
\end{subequations}

%

\subsection{Entanglement of axisymmetric states}

The optimal Schmidt number witness~\cite{Sanpera2001}
\begin{equation}
     \mathcal{W}\ =\ \ \frac{k-1}{d}\id_{d^2}\ -\ \ket{\Phi_d}\!\bra{\Phi_d}
\end{equation}
for Schmidt number $k$
can be used to detect the exact boundaries of the different SLOCC
classes (for $x>0$), that is, the regions of different Schmidt number
(see~Fig.~\ref{fig:triangle}).
While for $x<0$ the witness cannot be applied, one can check that
the projection $\mathbb{P}^{\mathrm{axi}}(\psi_0)$ of the product state
$\ket{\psi_0}=\frac12\left(\ket{1}+\ket{2}\right)\otimes\left(\ket{1}-\ket{2}\right)$
is the endpoint of the border for separable states. It is also easy 
to verify that above the line connecting this point with the separable 
state at
$(x=0,y=\sqrt{\frac{d-1}{d}})$ the negativity becomes nonzero, but does
not exceed 1. The state at the upper left corner is a state of
at most $r=2$ since it is the projection of 
$\oost\left(\ket{11}-\ket{22}\right)$, therefore the entangled states for $x<0$ must have $r=2$.

    The states with Schmidt number $\leqq k$
    belong to convex sets $S_k$ and form a hierarchy
    $S_1\subset S_2\subset\ldots\subset S_d$.
    Schmidt number $k=1$ corresponds
    to separable states (see Fig.~\ref{fig:triangle}).
Notably, all the boundaries are represented by straight lines.
In fact, this is a hint that the bipartite case---even for large $d$---is 
more treatable
than the multipartite case where also for highly symmetric families
of states the borders between entanglement classes are 
complicated~(cf.~\cite{Eltschka2012}). 

Let us now consider the entanglement
quantitatively. The formula for the negativity~\eqref{eq:defneg} is readily
applied to $\rho^{\mathrm{axi}}$ and gives 
\begin{align}
   & \mathcal{N}(\rho^{\mathrm{axi}}(x,y))  \nonumber\\
   &\;\;\;\;\;\;\ =\  \max  \left\{ 0 ,  \right.
       \left.\frac{1}{2}\left[ \sqrt{d(d-1)}|x|+\sqrt{d-1}y-\frac{d-1}{d}
                                             \right] \right\}\ .
\label{eq:axiN}
\end{align}
On the other hand, by using the lower bound (\ref{eq:huber2}) 
for the concurrence we obtain 

\begin{align}
    C(\rho^{\mathrm{axi}} & (x,y))  \geqq  \max\left\{ 0 ,  \right.
\nonumber\\
           & \left.  \!\! \sqrt{\frac{2}{d(d-1)}}\left[ \sqrt{d(d-1)}|x|+\sqrt{d-1}y-\frac{d-1}{d}
                                             \right]
                       \right\} .
\label{eq:axiC}
\end{align}
%
%
%
\begin{figure}[bt]
  \centering
 \includegraphics[width=.97\linewidth]{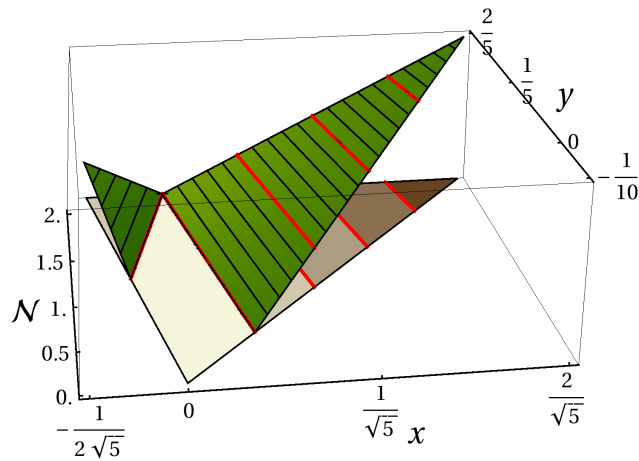}
  \caption{(color online). Negativity for the axisymmetric states with $d=5$
           according to Eq.~\eqref{eq:axiN}.
           The results for the concurrence $C(\rho^{\mathrm{axi}})$
           are qualitatively identical,
           the only difference is a scaling factor
           $\frac12 \sqrt{\frac{2d}{d-1}}$.
           The solid red lines indicate the borders between
           entanglement classes. Note that these are lines of
           constant fidelity, thus providing a nice illustration
           for the estimate in Eq.~\eqref{eq:FEFconc}.
    }
\label{fig:3d}
\end{figure}

%
%
These results (see Fig.~\ref{fig:3d}) are remarkable for several reasons. Both
concurrence and fidelity estimates depend linearly on $|x|$ and $y$
coordinates. Since Eqs.~\eqref{eq:axiN} and~\eqref{eq:axiC} include
the exact values for the pure state $\Phi_d$, and their graphs are planes,
there cannot be a larger convex function containing $\mathcal{N}(\Phi_d)$
and $C(\Phi_d)$ than these graphs.
Hence, the formulas~\eqref{eq:axiN} and~\eqref{eq:axiC} 
represent the exact solutions for the convex-roof extended negativity
and the concurrence of axisymmetric states, respectively.
Moreover, it follows that all PPT axisymmetric states are separable.
We mention that the concurrence result restricted to the isotropic states was
found in Ref.~\cite{Rungta2003}.

Another immediate consequence is that the integer part of 
$\left(2\mathcal{N}(\rho^{\mathrm{axi}})+1\right)$ changes by 1 
whenever a border between SLOCC classes, i.e., Schmidt numbers for 
$\rho^{\mathrm{axi}}$, is crossed. That is, 
for axisymmetric states  Eq.~\eqref{eq:negdimbound} can
be written 
\begin{equation}
   r(\rho^{\mathrm{axi}})\ =\ \lceil 2\mathcal{N}(\rho^{\mathrm{axi}})\rceil+1
\end{equation}
with the ceiling function $\lceil x \rceil$ denoting the smallest integer
greater than or equal to $x$. 

Concluding this section
we discuss yet another procedure to determine a lower 
bound for the convex-roof extended negativity and the concurrence
of arbitrary $d\times d$ states $\rho$. As we have mentioned before,
the symmetrization Eqs.~\eqref{eq:projaxi}--\eqref{eq:symmmatels} 
does not increase the entanglement 
\begin{align}
           \mathcal{N}^{\mathrm{CREN}}
                                      \left(
               \mathbb{P}^{\mathrm{axi}}(\rho)
                                      \right)\  & \leqq\ \
           \mathcal{N}^{\mathrm{CREN}}(\rho)\ \ ,\nonumber
\\
           C\left(\mathbb{P}^{\mathrm{axi}}(\rho)\right)\ & \leqq\ 
           C(\rho)
\end{align}
so that after symmetrizing $\rho$ we can simply read off the 
value for $\mathcal{N}^{\mathrm{CREN}}
                                      \left(
               \mathbb{P}^{\mathrm{axi}}(\rho)
                                      \right)$ 
or $C
                                      \left(
               \mathbb{P}^{\mathrm{axi}}(\rho)
                                      \right)$ 
from Fig.~\ref{fig:3d}. As discussed in Sec.~\ref{sec:lowerbounds} we can maximize
the state $\rho$ over local unitaries before projecting it and 
thus obtain an optimized lower bound. 

It is interesting to note
that for the concurrence this bound coincides with the one obtained
from Eq.~\eqref{eq:FEFconc}. The latter can be regarded as the result
of a projection of the optimized state onto the isotropic states. 
Thus we see that
one does not lose entanglement information projecting directly onto
the isotropic rather than the axisymmetric states. This is a
direct consequence of the fact that essential entanglement-related information
of a bipartite state is contained in its fidelity with the maximally entangled
state $\Phi_d$, i.e., the fully entangled fraction.

\section{Conclusion}

We have pointed out and made explicit that not only negativity,
but also concurrence is closely related to the partial transposition
of a $d\times d$ density matrix. In fact, both measures may be
understood and derived as quantifiers for the violation of the
PPT criterion in pure states. We have discussed that both negativity and
concurrence quantify the Schmidt rank of a pure state, however, in different
mathematical ways which hints at the fact that they quantify different
resources. Finally we have shown that, while the concurrence
equals the $\ell_2$ norm of the concurrence vector of a pure state, 
the negativity  is in general larger than the $\ell_1$ norm of
the concurrence vector. The negativity equals that $\ell_1$ norm 
if the pure state is written in the 
Schmidt decomposition.

These relations between negativity and concurrence lead to
various estimates for those measures (as well as for the 
Schmidt number) in mixed states. A particularly nice result
is that the negativity represents a direct lower bound to the Schmidt number
of a state, Eq.~\eqref{eq:negdimbound}. In the last section, we 
have provided an extensive discussion of the axisymmetric states,
a nontrivial family of highly symmetric $d\times d$ states for which
all the entanglement properties studied in this article 
can be calculated exactly.

%
%
%

\section*{Acknowledgements}
This work was funded by the German Research Foundation within 
SPP 1386 (C.E.), by Basque Government grant IT-472-10,
MINECO grants FIS2012-36673-C03-01  and  FIS2012-36673-C03-03, and UPV/EHU program UFI 11/55  
(J.S.\ and G.T.), the EU (ERC Starting Grant GEDENTQOPT, CHIST-ERA QUASAR), and 
National Research Fund of Hungary OTKA (Contract No. K83858).
C.E.\ and J.S.\ thank  M.\ Huber and G.\ Sent{\'i}s 
for stimulating discussions, 
and J.\ Fabian and K.\ Richter for their support.
%
%

%
%
%

\section*{references}

\end{document}